\pdfoutput=1

\documentclass [10pt,twocolumn]{IEEEtran}
\usepackage{algorithm,algorithmic,amsbsy,amsmath,amssymb,epsfig,bbm,mathrsfs,multirow,amsthm}
\usepackage{graphicx,caption,subcaption}
\newtheorem{proposition}{Proposition} 
\newtheorem{assumption} {Assumption}

\begin{document}
\title{Optimal Energy Allocation Policy for Wireless Networks in the Sky}
\author{Dinh Thai Hoang$^1$, Dusit Niyato$^1$ and Nguyen Tai Hung$^2$\\
$^1$ School of Computer Engineering, Nanyang Technological University (NTU), Singapore\\
$^2$ School of Electronics and Telecommunications, Hanoi University of Science and Technology, Vietnam\\}

\maketitle
\begin{abstract}
Google's Project Loon~\cite{LoonProject} was launched in 2013 with the aim of providing Internet access to rural and remote areas. In the Loon network, balloons travel around the Earth and bring access points to the users who cannot connect directly to the global wired Internet. The signals from the users will be transmitted through the balloon network to the base stations connected to the Internet service provider (ISP) on Earth. The process of transmitting and receiving data consume a certain amount of energy from the balloon, while the energy on balloons cannot be supplied by stable power source or by replacing batteries frequently. Instead, the balloons can harvest energy from natural energy sources, e.g., solar energy, or from radio frequency energy by equipping with appropriate circuits. However, such kinds of energy sources are often dynamic and thus how to use this energy efficiently is the main goal of this paper. In this paper, we study the optimal energy allocation problem for the balloons such that network performance is optimized and the revenue for service providers is maximized. We first formulate the stochastic optimization problem as a Markov decision process and then apply a learning algorithm based on simulation-based method to obtain optimal policies for the balloons. Numerical results obtained by extensive simulations clearly show the efficiency and convergence of the proposed learning algorithm. 
\end{abstract}

{\it Keywords-} Internet in the sky, Google Loon Project, Markov decision process.
 
\section{Introduction}
After more than 40 years of development Internet has created a revolution in communication for humans because it allows people to access and exchange information efficiently. Although Internet is highly accessible, approximately 60-70\% of people worldwide do not have the Internet reported by International Telecommunications Union~\cite{ITU_2013} in June 2013. This stems from a fact that many areas such as Africa, Asia, and Pacific, cannot offer Internet connections due to geographical and infrastructure issues. Therefore, the idea of providing Internet connections via wireless networks has become more and more popular. 

In wireless Internet, mobile users can connect to the Internet service provider (ISP) through base stations or access points. However, deployment of base stations for every location on the Earth seems to be impossible, e.g., oceans and mountains. Therefore, the idea of providing Internet from the sky was introduced. The early version is based on the satellites, which suffers from high cost and long transmission delay~\cite{Hu_Satellite_2001}. As a result, the cheaper and faster alternative, i.e., Google Loon project~\cite{LoonProject}, was proposed. In Loon project, access points will be placed on balloons flying at an altitude of about 20 km which is safe from bad weather and flights. The balloons will travel around the Earth and form a network of access points for Internet users in remote places. When receiving data from the user, the balloon will find the shortest route to transfer data to the nearest base station on the ground, which will be forwarded to an ISP.
\begin{figure}[htb]
\centering
\includegraphics[scale=0.56]{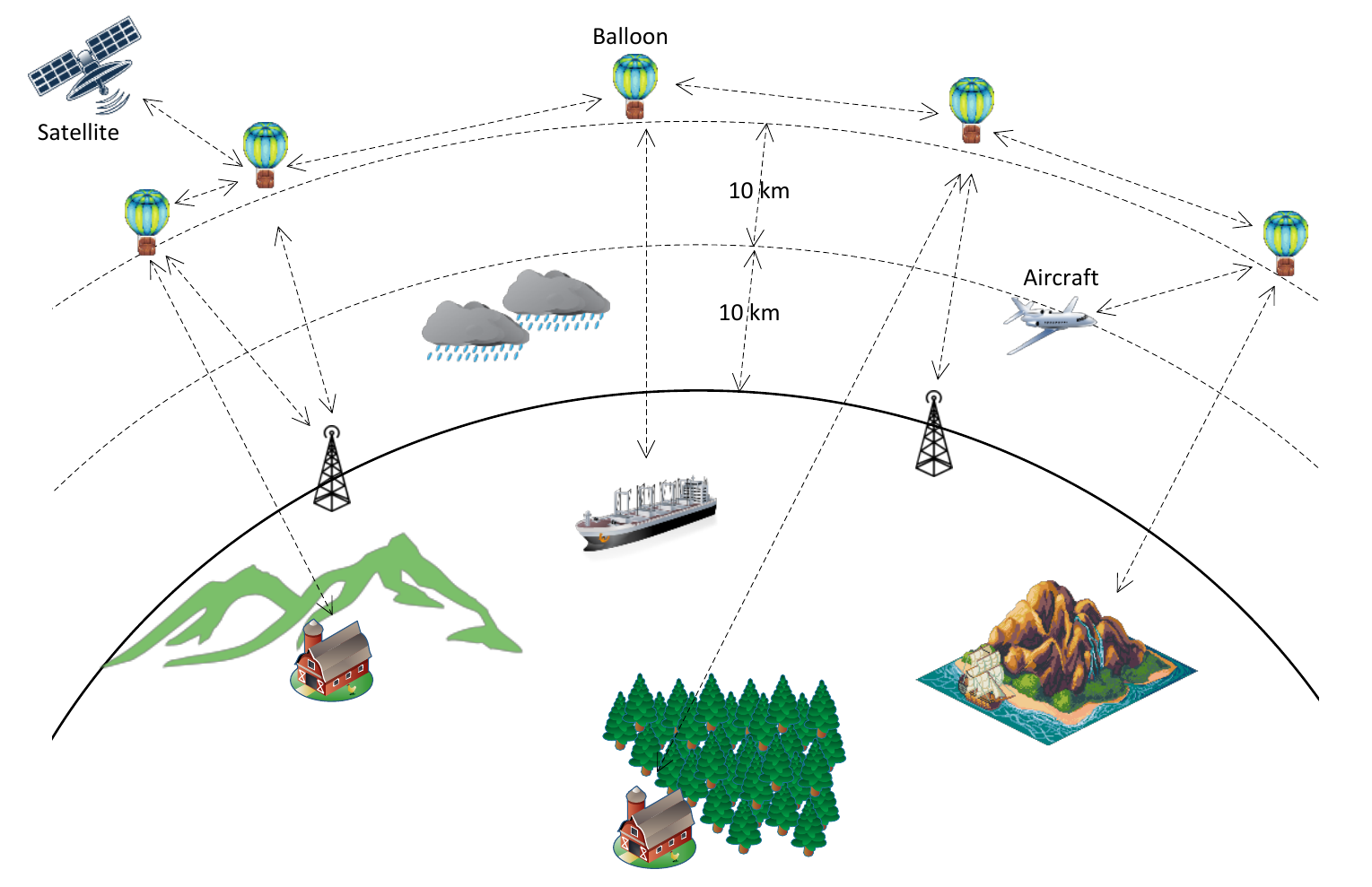}
\caption{The model of wireless network in the sky.}
\label{The system model}
\end{figure}

Data transmission and reception by the access points on the balloon consume energy, which requires continuous supply to sustain the operations. Only viable energy sources for the balloon are through energy harvesting such as solar energy or radio frequency (RF). The harvested energy will be stored in the energy storage of the access points and it will be used for data transmission and reception. However, batteries equipped on balloons are often limited by size and energy harvested from solar or RF is random. Moreover, the balloons may have to serve data transfer requests from different types of users, e.g., from other balloons, on the ground, satellites, or aircrafts, with different quality of service (QoS) requirements. Therefore, energy management for the balloons is an important issue.

In this paper, we aim to find an optimal admission control policy for the access points deployed on the balloons. The goal is to ensure high energy efficiency while maximizing profit of service providers. We formulate a Markov decision process (MDP) for the energy allocation optimization problem. To obtain the optimal policy, we apply a learning algorithm based on the policy gradient method and simulation-based method. The proposed learning algorithm not only avoids the curse of dimensionality problem caused by the explosion of state and action spaces, but also eliminates the need for complete knowledge about the model, which may not be possible to have from an unpredictable environment. Numerical results show the convergence as well as the efficiency of the proposed learning algorithm.

\section{System Model}
\begin{figure}[htb]
\centering
\includegraphics[scale=0.8]{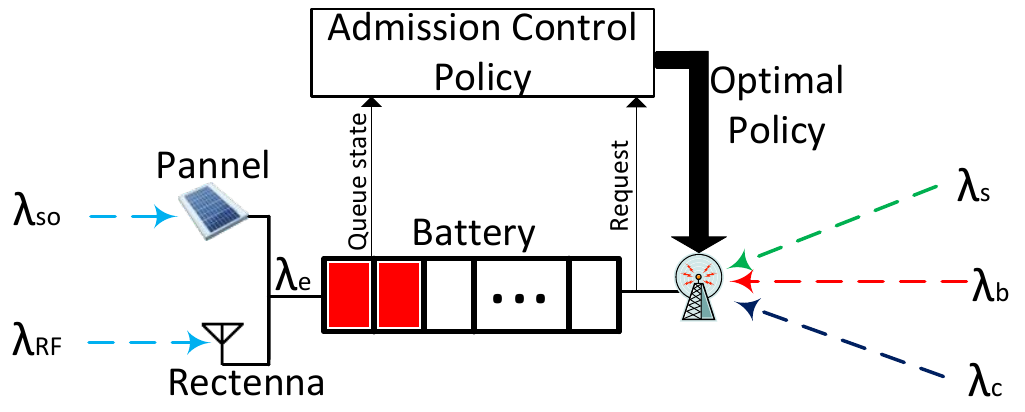}
\caption{The system model.}
\label{The system model 2}
\end{figure}
We consider energy management problem for a balloon with an access point as shown in Fig.~\ref{The system model 2}. Specifically, the access point is to receive and transmit data from users. The access point has a battery to store energy harvested from solar and RF. When a request is sent to the access point, it will check the amount of energy remaining in the battery and apply an admission control policy by deciding whether the request will be accepted or rejected. If the request is accepted, a certain amount of energy from the battery will be used to receive data and transmit data to the next hop. Different requests have different QoS requirements. Therefore, we divide requests into three classes, i.e., requests from other balloons (i.e., class-1), requests from users on the ground (class-2), and requests from satellites or aircrafts (class-3). The arrival processes of requests from class-1, class-2, and class-3 are assumed to follow the Poisson distribution with mean rates $\lambda_b$, $\lambda_c$, and $\lambda_s$, respectively. When a request is accepted, the balloon will receive an immediate reward (i.e., revenue). The immediate revenues for accepting requests from class-1, class-2, and class-3 are $r_b$, $r_c$, and $r_s$, respectively.

In the Loon network, balloons are assumed to be equipped with solar panels~\cite{Raghunathan2005} for harvesting energy from sunlight. Additionally, we assume that the balloons can be equipped with a rectenna~\cite{Georgiadis2010} to harvest energy from RF waves. We assume that the energy arrival from both solar panel and RF rectenna follows the Poisson distribution with mean rates $\lambda_{e}$, and the successful energy harvesting probability is $p_{e}^{su}$. The maximum capacity of the battery is $E$. At each time epoch, when the access point receives a request, it will consult with the admission control policy. The admission control policy will determine a decision to accept or reject the request based on the request' class and the current energy level in the battery. In this paper, we are interested in maximizing the profit for the service provider in the terms of average reward for the balloon. 

\section{Problem Formulation}
\label{sec:problem_formulate}
In this Section, we will formulate the optimization problem as Markov decision process (MDP) and study a learning algorithm to obtain the optimal policy for balloons.

\subsection{MDP Framework}
An MDP is defined by a tuple of $<\mathcal{S}, \mathcal{A}, P, R>$ where $\mathcal{S}$ is a state space, $\mathcal{A}$ is an action space, $P$ is a transition probability function, and $R$ is a reward function. The state space $\mathcal{S}$ of our admission control for the access point on a balloon is defined as follows:
\begin{equation}
	{\mathcal{S}} \triangleq \Big\{	{\mathbf{s}=(\mathbf{e}, \mathbf{x})}: {\mathbf{e}} \in \mathcal{E}; \mathbf{x} \in \mathcal{X} \Big\}	,
\end{equation}
where $\mathcal{E}=\{0,1,\ldots,E\}$ is the energy state space whose elements represent energy levels in the battery. $\mathcal{X}=\{x_b,x_c,x_s,x_e\}$ is a set of events in which $x_b$, $x_c$, and $x_s$ are the events when a request is from another access point, a user, and a satellite, respectively. $x_e$ is an event for energy arrival.

When the system is at state $\mathbf{e}$, if an event $\mathbf{x}$ happens, an accept/reject decision must be made. Thus, we can define the action space as follows:
\begin{equation}
	{\mathcal{A}} \triangleq \Big\{	\mathbf{a}: {\mathbf{a}} \in \{0,1\} \Big\}	,
\end{equation}
where 
\begin{displaymath}
\mathbf{a} = \left\{ 
\begin{array}{ll}
1, & \text{if the arriving request is accepted} \\
0, & \text{if the arriving request is rejected}
\end{array}
\right.
\end{displaymath}

To derive the transition probability function $P(\mathbf{s},\mathbf{a},\mathbf{s'})$, we consider a discrete time system by using uniformization technique~\cite{Gallager95} with a uniformization parameter $u$ obtained as follows:
\begin{equation}
u =\lambda_b + \lambda_c +\lambda_s + \lambda_{e}	.
\end{equation}
Based on the uniformization parameter $u$, we can determine the probabilities of the events as follows. In the event $\mathbf{e}$, the probabilities of a request arriving from other balloons, a user, and a satellite/an aircraft are $\lambda_b/u$, $\lambda_c/u$, $\lambda_s/u$, respectively. The probability of energy arrival is $\lambda_e/u$. Then, we can derive the transition probability matrix for the system. However, to do so, we need to know the environment parameters, e.g., successful energy harvesting probability, requests' arrival rates. These parameters are generally not known in advance and building the model with complete information may not be possible. Therefore, we apply a learning algorithm based on simulation-based method~\cite{Gosavi2003}. The main idea of the simulation-based method is based on a ``simulator'' that can simulate the environment by generating environment parameters (e.g., a successful energy harvesting probability and arrival rates). Then we use the parameters from simulations to derive the admission control policy for the access point. Based on the simulation-based method, the transition probability function can be defined as follows:
\begin{equation}
P(\mathbf{s},\mathbf{a},\mathbf{s'}) = p_{env} p(\mathbf{s}) p(\mathbf{a})	,
\end{equation}
where $p_{env}$ is environment parameter (e.g., the successful energy harvesting probability), $p(\mathbf{s})$ is the probability that the system is at state $\mathbf{s}$, and $p(\mathbf{a})$ is the probability that action $\mathbf{a}$ is taken. 

When there is a request $\mathbf{x}$ arriving at the access point, if the request is accepted, the balloon will receive an immediate reward $r_x$ corresponding to the type of the request. Otherwise, the access point gains nothing, i.e., 
\begin{displaymath}
R(\mathbf{s}, \mathbf{a}) = \left\{ 
\begin{array}{ll}
r_{x}, & \text{if} \phantom{1} \mathbf{x} \in \{x_b,x_c,x_s\}, \mathbf{a}=1 \phantom{1} \text{and} \phantom{1} \mathbf{e}+\mathbf{x} \in \mathcal{E} \\
0, & \text{otherwise.}
\end{array}
\right.
\end{displaymath}
Note that, for the case $\mathbf{x} = x_e$, the access point will always receive energy if the battery is not full. However, there is no reward for such an action.

\subsection{MDP with Parameterization}

We consider a parameterized randomized policy~\cite{Marbach2001, Olivier2007, Baxter2001}. With the parameterized randomized policy, when there is a request arriving at the access point, the request will be accepted with probability defined as follows:
\begin{equation}
\label{eq:parameterized policy}
\mu_{\Theta}(\mathbf{s},\mathbf{a}) = \frac {1} {1+\exp(1.5(\theta_{x}-\mathbf{e}))}	,
\end{equation}
where $\Theta$ is the parameter vector of the learning algorithm, $\mathbf{e}$ is the current energy level of the battery and $\theta_{x}$ is the parameter for requests from event $\mathbf{x}$. Additionally, the parameterized randomized policy $\mu_{\Theta}(\mathbf{s},\mathbf{a})$ must not be negative and meet the following condition,
\begin{equation}
\label{for:randomized_policy}
\sum_{\mathbf{a} \in \mathcal{A}} \mu_{\Theta}(\mathbf{s}, \mathbf{a}) = 1	.
\end{equation}

With the randomized parameterized policy, the transition probability function will be parameterized as follows:
\begin{equation}
P_{\Theta}(\mathbf{s},\mathbf{s'}) = \sum_{\mathbf{a} \in \mathcal{A}} \mu_{\Theta}(\mathbf{s}, \mathbf{a}) P(\mathbf{s},\mathbf{a},\mathbf{s'})	.
\end{equation}
Similarly, we can parameterize the immediate reward function as follows: 
\begin{equation}
\label{eq:parameterized_reward}
R_{\Theta}(\mathbf{s}) = \sum_{\mathbf{a} \in \mathcal{A}} \mu_{\Theta}(\mathbf{s}, \mathbf{a})R(\mathbf{s}, \mathbf{a})	.
\end{equation}

We aim to maximized the average reward under randomized parameterized policy denoted by $\psi(\Theta)$ and it can be defined as follows:
\begin{equation}
\psi(\Theta) = \lim_{t\rightarrow \infty} \frac {1}{t} E_{\Theta} \Bigg[ \sum_{k=0}^{t} R_{\Theta}(\mathbf{s}_{k}) \Bigg]	,
\end{equation}
where $\mathbf{s}_k$ is the system state at step $k$ and $E_{\Theta}[\cdot]$ is the expected reward of the system.

We then make some assumptions as follows:
\begin{assumption}
\label{recurrent_state}
The Markov chain corresponding to every $P \in \overline{\mathcal{P}}$ is aperiodic. Furthermore, there exists a state $\mathbf{s}^{*}$ which is recurrent for every of such Markov chain. 
\end{assumption}

\begin{assumption}
\label{derivatives}
For every state $\mathbf{s,s'} \in \mathcal{S}$, the functions $P_{\Theta}(\mathbf{s,s'})$ and $R_{\Theta}(\mathbf{s})$ are bounded, twice differentiable, and have bounded first and second derivatives. 
\end{assumption}

Assumption~\ref{recurrent_state} implies that the system has a Markov property and Assumption~\ref{derivatives} guarantees that the transition probability function and the average reward function depend smoothly on the parameter vector $\Theta$ after they are parameterized by $\Theta$. Assumption~\ref{derivatives} is necessary when we use the policy gradient method to adjust vector $\Theta$. Under Assumption~\ref{derivatives}, the average reward $\psi({\Theta})$ is well defined for every $\Theta$ and does not depend on an initial state. Furthermore, we have the following balance equations:
\begin{eqnarray}
\lefteqn{ \sum_{\mathbf{s}=1}^{S} \pi_{\Theta}(\mathbf{s}) P_{\Theta}(\mathbf{s,s'}) = \pi_{\Theta}(\mathbf{s'}), }\nonumber\\
& & {} \sum_{\mathbf{s}=1}^{S} \pi_{\Theta}(\mathbf{s})  = 1  {}	,
\end{eqnarray}
where $\pi_{\Theta}(\mathbf{s})$ is steady state probability at state $\mathbf{s}$ under the parameter vector, and thus the average reward function can be also defined as follows:
\begin{equation}
\psi(\Theta) = \sum_{\mathbf{s}} \pi_{\Theta}(\mathbf{s}) R_{\Theta}(\mathbf{s})	.
\end{equation}

\subsection{Policy Gradient Method}
We now can apply the policy gradient method~\cite{Bertsekas1995} to update for the parameter vector $\Theta$ as follows:
\begin{equation}
\label{eq:policy_gradient_method}
\Theta_{k+1} = \Theta_{k} + \gamma_{k} \nabla \psi(\Theta_{k}) 
\end{equation}
where $\gamma_k$ is a step size parameter. In the policy gradient method, we start with an initial parameter vector $\Theta_0$, and then the parameter vector $\Theta$ will be updated iteratively based on (\ref{eq:policy_gradient_method}). Under Assumption~\ref{derivatives} and an appropriate step size, it was proved in~\cite{Bertsekas1995} that, $\lim_{k \rightarrow \infty} \nabla \psi(\Theta_{k}) = 0$. That is, the average reward $\psi(\Theta_k)$ converges almost surely. 

We now propose Proposition~\ref{prop:policy_gradient} to calculate the gradient for the average reward $\psi(\Theta)$.
\begin{proposition}
\label{prop:policy_gradient}
Let Assumption~\ref{recurrent_state} and Assumption~\ref{derivatives} hold, then \\
\begin{equation}
\label{eq:gradient_average_reward}
\nabla \psi(\Theta) = \sum_{\mathbf{s} \in \mathcal{S}} \pi_{s}(\Theta) \Big(\nabla R_{\Theta}(\mathbf{s}) + \sum_{\mathbf{s'} \in \mathcal{S}} \nabla P_{\Theta}(\mathbf{s,s'}) \mathbf{d}_{\Theta}(\mathbf{s'})  \Big) .
\end{equation}
\end{proposition}
$\mathbf{d}_{\Theta}(\mathbf{s'})$ is the differential reward at state $\mathbf{s}$ and it can be defined as follows:
\begin{equation}
d_{\Theta}(\mathbf{s}) = E_{\Theta} \Bigg[ \sum_{k=0}^{T-1} (R_{\Theta}(\mathbf{s}_{k}) - \psi(\Theta) | \mathbf{s_{0}} = \mathbf{s} \Bigg]	,
\end{equation} 
where $T=\min\{k>0|\mathbf{s}_k=\mathbf{s}^*\}$ is the first future time that the state $\mathbf{s}^*$ is visited. Because of limited space, the proof of Proposition~\ref{prop:policy_gradient} can be found in~\cite{Marbach2001}.

\subsection{Simulation-based learning algorithm}
We update the parameter vector $\Theta$ iteratively based on (\ref{eq:policy_gradient_method}) with the value of the gradient of average reward calculated from Proposition~\ref{prop:policy_gradient}. However, it is not easy to calculate the terms in (\ref{eq:gradient_average_reward}). Additionally, when the state space and action space are large, it is intractable to calculate exactly the value of the gradient of the average reward function. Therefore, in this paper, we consider the approach that can estimate the gradient of the average reward function and then the parameter vector $\Theta$ can be adjusted in an online manner. 

From~(\ref{for:randomized_policy}), we have $\sum_{\mathbf{a} \in \mathcal{A}} \mu_{\Theta}(\mathbf{s}, \mathbf{a}) = 1$, so we derive $\sum_{\mathbf{a} \in \mathcal{A}} \nabla \mu_{\Theta}(\mathbf{s}, \mathbf{a}) = 0$. From~(\ref{eq:parameterized_reward}), we have:
\begin{equation}
\begin{aligned}
\nabla R_{\Theta}(\mathbf{s})& = \sum_{\mathbf{a} \in \mathcal{A}} \nabla \mu_{\Theta}(\mathbf{s}, \mathbf{a})R(\mathbf{s}, \mathbf{a})\\
&=\sum_{\mathbf{a} \in \mathcal{A}} \nabla \mu_{\Theta}(\mathbf{s}, \mathbf{a}) \big(R(\mathbf{s}, \mathbf{a}) - \psi(\Theta)\big).
\end{aligned}
\end{equation}
This is from the fact that $\sum_{\mathbf{a} \in \mathcal{A}} \nabla \mu_{\Theta}(\mathbf{s}, \mathbf{a}) = 0$. Moreover, we have 
\begin{equation}
\sum_{\mathbf{s'} \in \mathcal{S}}\nabla P_{\mathbf{s,s'}}(\Theta) \mathbf{d}(\mathbf{s'},\Theta) = \sum_{\mathbf{s'} \in \mathcal{S}} \sum_{\mathbf{a} \in \mathcal{A}} \nabla \mu_{\Theta}(\mathbf{s}, \mathbf{a}) P_{\mathbf{s,s'}}(\mathbf{a}) \mathbf{d}(\mathbf{s'},\Theta).
\end{equation}

Therefore, along with Proposition~\ref{prop:policy_gradient}, we derive the following results:
\begin{equation}
\begin{aligned}
  \nabla \psi(\Theta)&=\sum_{\mathbf{s} \in \mathcal{S}} \pi_{\Theta}(\mathbf{s}) \Big(\nabla R_{\Theta}(\mathbf{s}) + \sum_{\mathbf{s'} \in \mathcal{S}}\nabla P_{\Theta}(\mathbf{s,s'}) \mathbf{d}_{\Theta}(\mathbf{s'}) \Big) \\
      &=\sum_{\mathbf{s} \in \mathcal{S}} \pi_{\Theta}(\mathbf{s}) \Big(\sum_{\mathbf{a} \in \mathcal{A}} \nabla \mu_{\Theta}(\mathbf{s}, \mathbf{a}) (R(\mathbf{s}, \mathbf{a}) - \psi(\Theta)) +\\
	&\phantom{5}+ \sum_{\mathbf{s'} \in \mathcal{S}}\sum_{\mathbf{a} \in \mathcal{A}} \nabla \mu_{\Theta}(\mathbf{s}, \mathbf{a}) P (\mathbf{s,a,s'}) \mathbf{d}_{\Theta}(\mathbf{s'}) \Big) \\
	&= \sum_{\mathbf{s} \in \mathcal{S}} \pi_{\Theta}(\mathbf{s}) \sum_{\mathbf{a} \in \mathcal{A}} \nabla \mu_{\Theta}(\mathbf{s}, \mathbf{a})  \\
	&\phantom{5}\times\Big(\big(R(\mathbf{s}, \mathbf{a}) - \psi(\Theta)\big) + P (\mathbf{s,a,s'}) \mathbf{d}_{\Theta}(\mathbf{s'}) \Big) \\ 
	&= \sum_{\mathbf{s} \in \mathcal{S}} \sum_{\mathbf{a} \in \mathcal{A}} \pi_{\Theta}(\mathbf{s}) \nabla \mu_{\Theta}(\mathbf{s}, \mathbf{a}) q_{\Theta}(\mathbf{s},\mathbf{a}),
\end{aligned}
\end{equation}
where 
\begin{equation}
\begin{aligned}
q_{\Theta}(\mathbf{s},\mathbf{a}) &= \Big(R(\mathbf{s}, \mathbf{a}) - \psi(\Theta)\Big) + \sum_{\mathbf{s'} \in \mathcal{S}} P (\mathbf{s,a,s'}) \mathbf{d}_{\Theta}(\mathbf{s'}) \\
&= E_{\Theta} \Bigg[\sum_{k=0}^{T-1}\big(R(\mathbf{s}, \mathbf{a}) - \psi(\Theta) \big) | \mathbf{s}_{0} = \mathbf{s}, \mathbf{a}_{0}=\mathbf{a} \Bigg].
\end{aligned}
\end{equation}

Here again $T$ is the first future time that the current state $\mathbf{s}^{*}$ is visited. $q_{\Theta}(\mathbf{s},\mathbf{a})$ can be interpreted as the differential reward if action $\mathbf{a}$ is taken based on policy $\mu_{\Theta}$ at state $\mathbf{s}$. We need to note that $d_{\Theta}(\mathbf{s})$ is the cost at state $\mathbf{s}$ and it is different from the different cost at state $\mathbf{s}$ under action $\mathbf{a}$, i.e., $q_{\Theta}(\mathbf{s},\mathbf{a})$. Then, we present Algorithm~\ref{algorithm0} to update the parameter vector $\Theta$ at the visits to the recurrent state $\mathbf{s}^{*}$. 

\begin{algorithm}
\caption{Algorithm to update parameter vector $\Theta$ at visits to the recurrent state $\mathbf{s}^{*}$}
\label{algorithm0}
At the time $T_{m+1}$ of the $(m+1)$th visit to state $\mathbf{s}^{*}$, we update the parameter vector $\Theta$ and the estimated average reward $\widetilde{\psi}$ as follows: \\
\begin{displaymath}
\Theta_{m+1} = \Theta_{m} + \gamma_{m}F_{m}(\Theta_{m},\widetilde{\psi}_{m}),
\end{displaymath}
\begin{displaymath}
\widetilde{\psi}_{m+1} = \widetilde{\psi}_{m} + \eta\gamma_{m}\sum_{n=t_{m}}^{t_{m+1}-1}(R(\mathbf{s}_{n}, \mathbf{a}_{n}) - \widetilde{\psi}_{m})
\end{displaymath}
where
\begin{displaymath}
F_{m}(\Theta_{m},\widetilde{\psi}_{m}) = \sum_{n=t_{m}}^{t_{m+1}-1} \widetilde{q}_{\Theta_{m}}(\mathbf{s}_{n},\mathbf{a}_{n}) \frac{\nabla \mu_{\Theta_{m}}(\mathbf{s}_{n},\mathbf{a}_{n})}{\mu_{\Theta_{m}}(\mathbf{s}_{n},\mathbf{a}_{n})},
\end{displaymath}
\begin{displaymath}
\widetilde{q}_{\Theta_{m}}(\mathbf{s}_{n},\mathbf{a}_{n}) = \sum_{k=n}^{t_{m+1}-1}(R(\mathbf{s}_{k}, \mathbf{a}_{k}) - \widetilde{\psi}_{m}).
\end{displaymath}
\end{algorithm}

In Algorithm~\ref{algorithm0}, $\eta$ is a positive scalar and $\gamma_{m}$ is a step size parameter. We derive the following convergence result for Algorithm~\ref{algorithm0}.

\begin{proposition}
\label{prop2}
Let Assumption~\ref{recurrent_state} and Assumption~\ref{derivatives} hold, and let ($\Theta_{m}$) be the sequence of parameter vectors generated by Algorithm~\ref{algorithm0} with a suitable step size parameter $\gamma$ satisfied Assumption~\ref{ass:step size}, then  $\psi({\Theta_{m}})$ converges and 
\begin{displaymath}
\lim_{m\rightarrow \infty} \nabla \psi(\Theta_{m}) = 0,
\end{displaymath}
with probability 1. 
\end{proposition}
The proof of the Proposition~\ref{prop2} can be found in~\cite{Marbach2001}.

\begin{assumption}
\label{ass:step size} 
The step size $\gamma_{m}$ is deterministic, nonnegative and satisfies the following condition,
\begin{displaymath}
\sum_{m=1}^{\infty}\gamma_{m} = \infty, \phantom{5} \sum_{m=1}^{\infty}\gamma_{m}^{2} < \infty	.
\end{displaymath}
\end{assumption}

In Algorithm~\ref{algorithm0}, to update the value of the parameter vector $\Theta$ at the next visit time to the state $\mathbf{s}^{*}$, we need to store all values of $\widetilde{q}_{\Theta_{m}}(\mathbf{s}_{n},\mathbf{a}_{n})$ and $\frac{\nabla \mu_{\Theta_{m}}(\mathbf{s}_{n},\mathbf{a}_{n})}{\mu_{\Theta_{m}}(\mathbf{s}_{n},\mathbf{a}_{n})}$ between two successive visits. However, this method could result in slow processing. Therefore, we modify Algorithm~\ref{algorithm0} to improve the efficiency. First, we rewrite $F_{m}(\Theta_{m},\widetilde{\psi}_{m})$ as follows:
\begin{equation}
\begin{aligned}
  F_{m}(\Theta_{m},\widetilde{\psi}_{m})&= \sum_{n=t_{m}}^{t_{m+1}-1} \widetilde{q}_{\Theta_{m}}(\mathbf{s}_{n},\mathbf{a}_{n}) \frac{\nabla \mu_{\Theta_{m}}(\mathbf{s}_{n},\mathbf{a}_{n})}{\mu_{\Theta_{m}}(\mathbf{s}_{n},\mathbf{a}_{n})} \\
      &= \sum_{n=t_{m}}^{t_{m+1}-1}\frac{\nabla \mu_{\Theta_{m}}(\mathbf{s}_{n},\mathbf{a}_{n})}{\mu_{\Theta_{m}}(\mathbf{s}_{n},\mathbf{a}_{n})} \sum_{k=n}^{t_{m+1}-1}\big(R(\mathbf{s}_{k}, \mathbf{a}_{k}) - \widetilde{\psi}_{m}\big) \\
	&= \sum_{n=t_{m}}^{t_{m+1}-1} \big(R(\mathbf{s}_{k}, \mathbf{a}_{k}) - \widetilde{\psi}_{m}\big) z_{k+1}, 
\end{aligned}
\end{equation}
where
\begin{displaymath}
z_{k+1} = \left\{ 
\begin{array}{ll}
\frac{\nabla \mu_{\Theta_{m}}(\mathbf{s}_{k},\mathbf{a}_{k})}{\mu_{\Theta_{m}}(\mathbf{s}_{k},\mathbf{a}_{k})}, & \text{if} \phantom{1} k = t_{m},\\
z_{k}+\frac{\nabla \mu_{\Theta_{m}}(\mathbf{s}_{k},\mathbf{a}_{k})}{\mu_{\Theta_{m}}(\mathbf{s}_{k},\mathbf{a}_{k})}, & k=t_{m}+1,\ldots,t_{m+1}-1. \\
\end{array}
\right.
\end{displaymath}

The detail of the algorithm can be expressed as in Algorithm~\ref{algorithm1}, where $\eta$ is a positive scalar and $\gamma_{k}$ is the step size of the algorithm. 

\begin{algorithm}
\caption{Algorithm to update $\Theta$ at every time step}
\label{algorithm1}
At a typical time $k$, the state is $\mathbf{s}_{k}$, and the values of $\Theta_{k}, z_{k}$, and $\widetilde{\psi}(\Theta_{k})$ are available from the previous iteration. We update $\Theta$ and $\widetilde{\psi}$ according to: \\
\begin{displaymath}
z_{k+1} = \left\{ 
\begin{array}{ll}
\frac{\nabla \mu_{\Theta_{k}}(\mathbf{s}_{k},\mathbf{a}_{k})}{\mu_{\Theta_{k}}(\mathbf{s}_{k},\mathbf{a}_{k})}, & \text{if} \phantom{1} \mathbf{s}_{k} = \mathbf{s*}\\
z_{k}+\frac{\nabla \mu_{\Theta_{k}}(\mathbf{s}_{k},\mathbf{a}_{k})}{\mu_{\Theta_{k}}(\mathbf{s}_{k},\mathbf{a}_{k})}, & \text{otherwise,} \\
\end{array}
\right.
\end{displaymath}
\begin{displaymath}
\Theta_{k+1} = \Theta_{k} + \gamma_{k}(R(\mathbf{s}_{k},\mathbf{a}_{k})-\widetilde{\psi}_{k})z_{k+1}, 
\end{displaymath}
\begin{displaymath}
\widetilde{\psi}_{k+1} = \widetilde{\psi}_{k} + \eta\gamma_{k}(R(\mathbf{s}_{k},\mathbf{a}_{k}) - \widetilde{\psi}_{k}).
\end{displaymath}
\end{algorithm}

\section{Numerical Results} 
\begin{figure*}[htb]
\centering
\begin{subfigure}{0.55\columnwidth}
\includegraphics[width=\columnwidth]{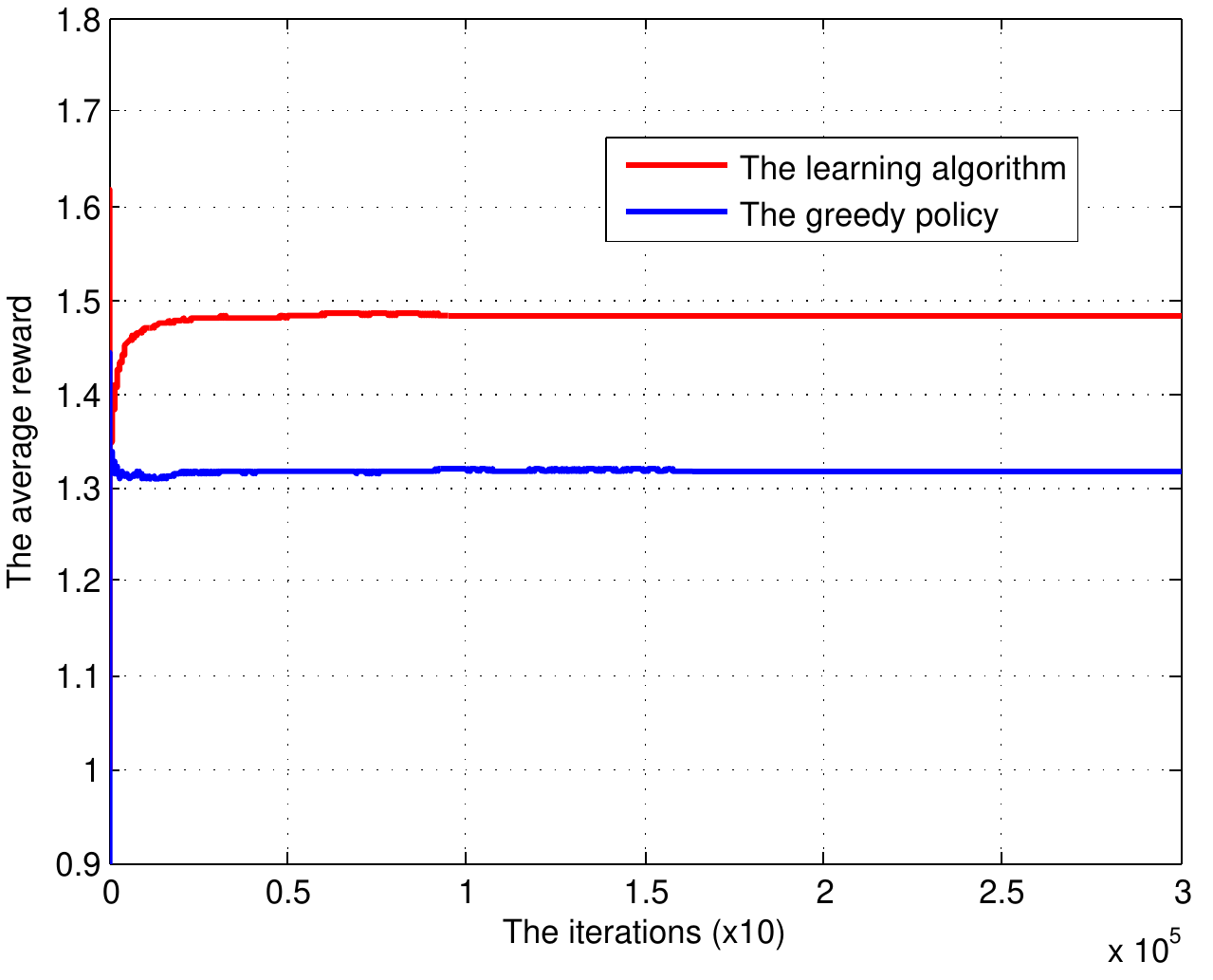}%
\caption{The average reward}%
\label{fig:sub_a}%
\end{subfigure}\hfill%
\begin{subfigure}{0.55\columnwidth}
\includegraphics[width=\columnwidth]{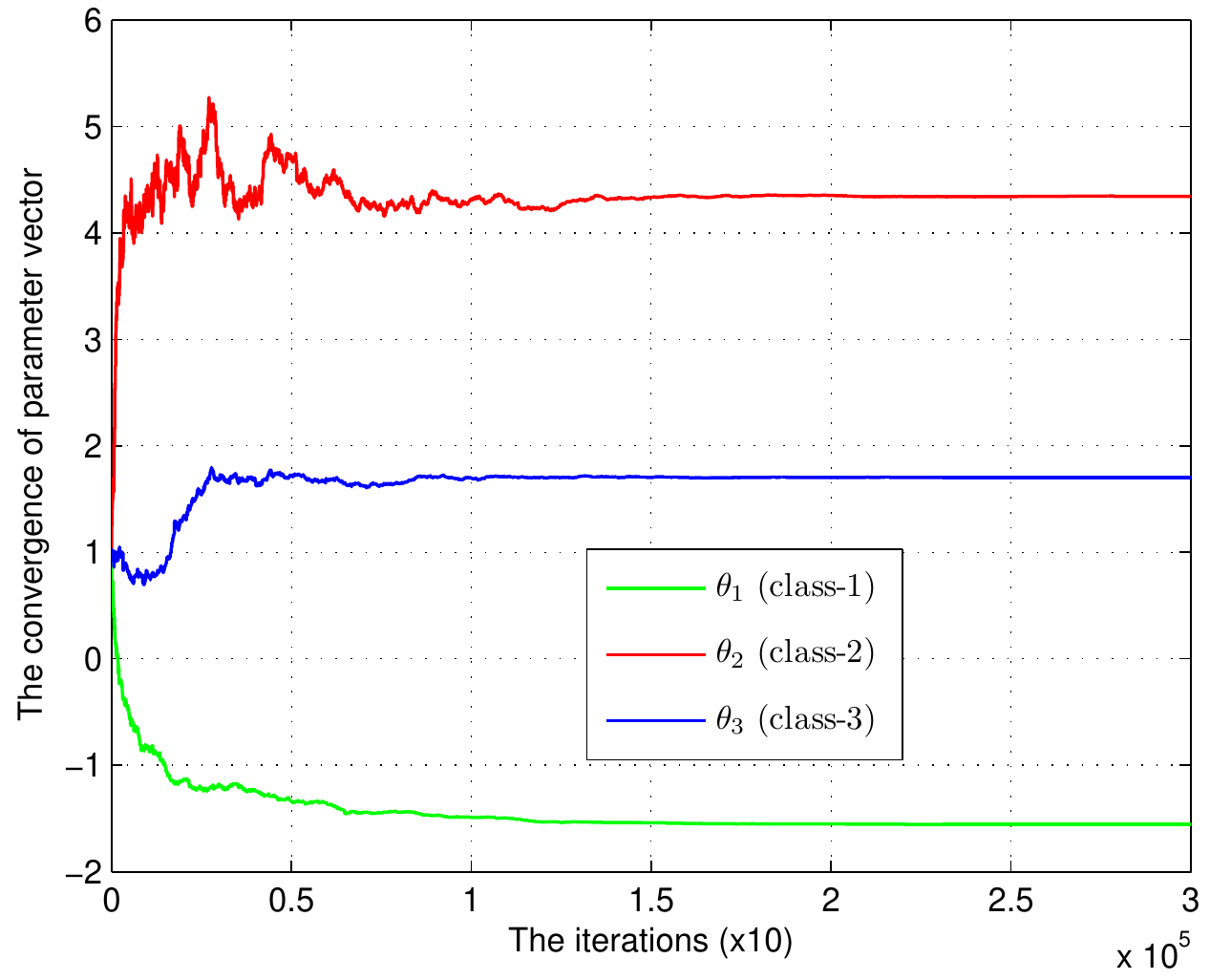}%
\caption{The parameter vector}%
\label{fig:sub_b}%
\end{subfigure}\hfill%
\begin{subfigure}{0.55\columnwidth}
\includegraphics[width=\columnwidth]{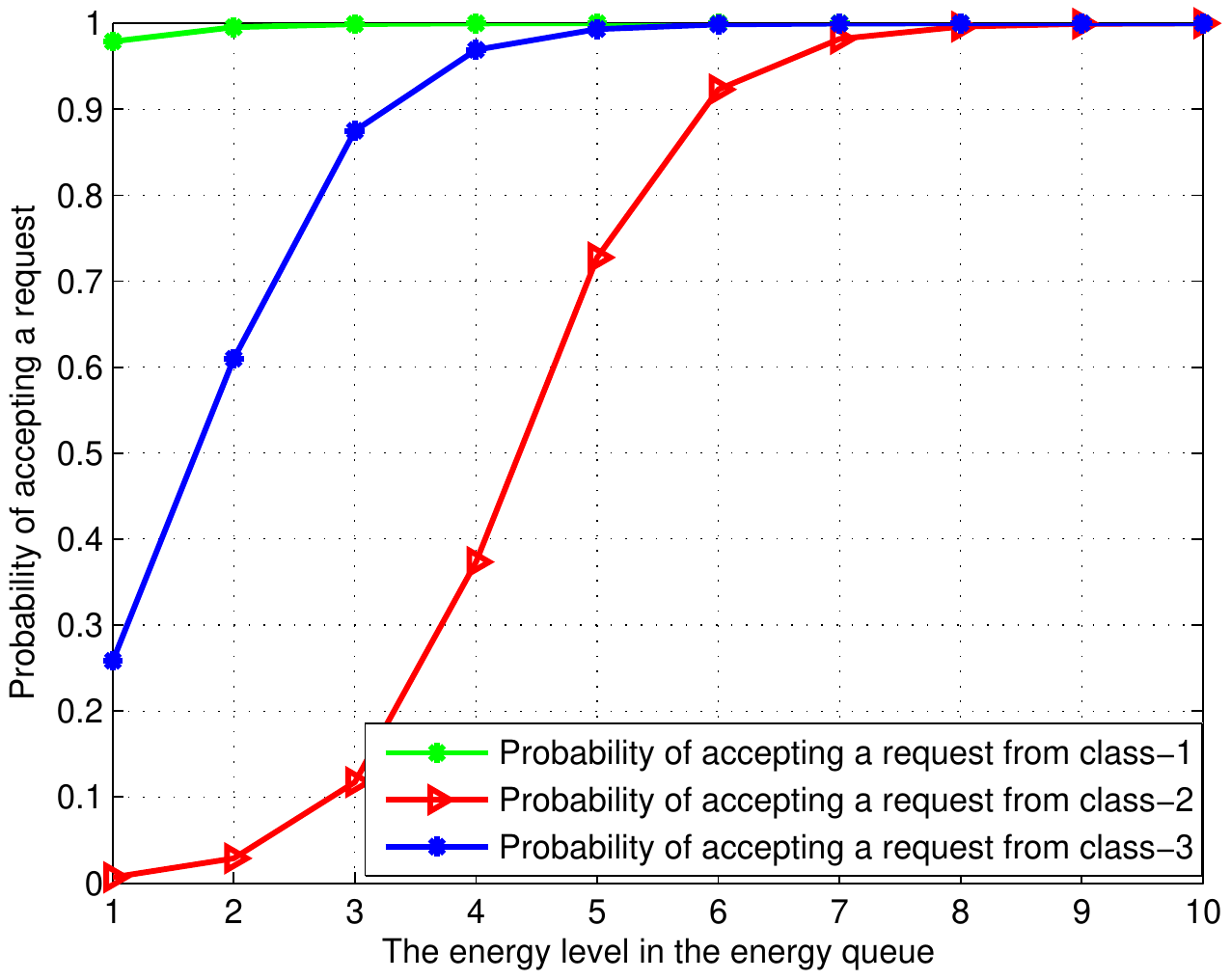}%
\caption{The probability of accepting requests from different classes of user}%
\label{fig:sub_c}%
\end{subfigure}%
\caption{The convergence and the policy of the learning algorithm.}
\label{fig:convergence_policy}
\end{figure*}
\subsection{Experiment Setup}
In this section, we perform simulations using MATLAB to evaluate the performance of the proposed learning algorithm. In the experiment, we consider the scenario as depicted in Fig.~\ref{The system model 2}. The maximum queue size is set at $10$ units. There are three classes of users, namely, class-1, class-2, and class-3, corresponding to requests from balloons, users on the ground, and satellites/aircrafts, respectively. The arrival rates of requests from users of class-1, class-2, and class-3, are $60$, $70$, and $10$ requests per hour, respectively. When a request is accepted, the access point will use one unit of energy from the battery to serve for the request (i.e., to receive data and transmit the data to the destination). Upon accepting requests, the access point receives the rewards of $5$, $2$, and $3$ monetary units for class-1, class-2, and class-3, respectively. The energy arrival rate is $110$ per hour and the successful energy harvesting probability is $0.9$. If the balloon harvest energy successfully and the battery is not full, the battery will increase one unit. For the learning algorithm, the initial parameter vector is set at $\Theta=(\theta_1, \theta_2, \theta_3)=(1,1,1)$, and the chosen initial estimated average reward is $0.7$.

\subsection{Numerical Results}
We first consider the convergence of the proposed learning algorithm (i.e., Algorithm~\ref{algorithm1}). Figures~\ref{fig:sub_a} and~\ref{fig:sub_b} show the convergence in the terms of the average reward and the parameter vector. In both figures, the proposed learning algorithm converges within around $5.10^5$-$10^6$ iterations. In Fig.~\ref{fig:sub_a}, we also compare the average rewards obtained by the learning algorithm and the greedy policy that always accepts requests. At the convergence points, the average reward obtained by the learning algorithm reaches approximately $1.48$ which $8.8$\% higher than that obtained by the greedy policy. 

In Fig.~\ref{fig:sub_b}, the parameter vector $\Theta$ converges to (-1.5577, 4.3448, 1.7029) for class-1, class-2, and class-3, respectively. Then, from the parameter vector obtained from the learning algorithm, we can determine the policy for the access point as shown in Fig.~\ref{fig:sub_c}. In Fig.~\ref{fig:sub_c}, the requests from other balloons will be always accepted, while the requests from users on the ground and satellites will only be accepted only when the energy level in the battery is high enough. In particular, the access point will accept the requests from a user on the ground and satellites when the energy level is higher than $5$ units and $2$ units, respectively. 

\begin{figure}[htb]
\centering
\includegraphics[scale=0.43]{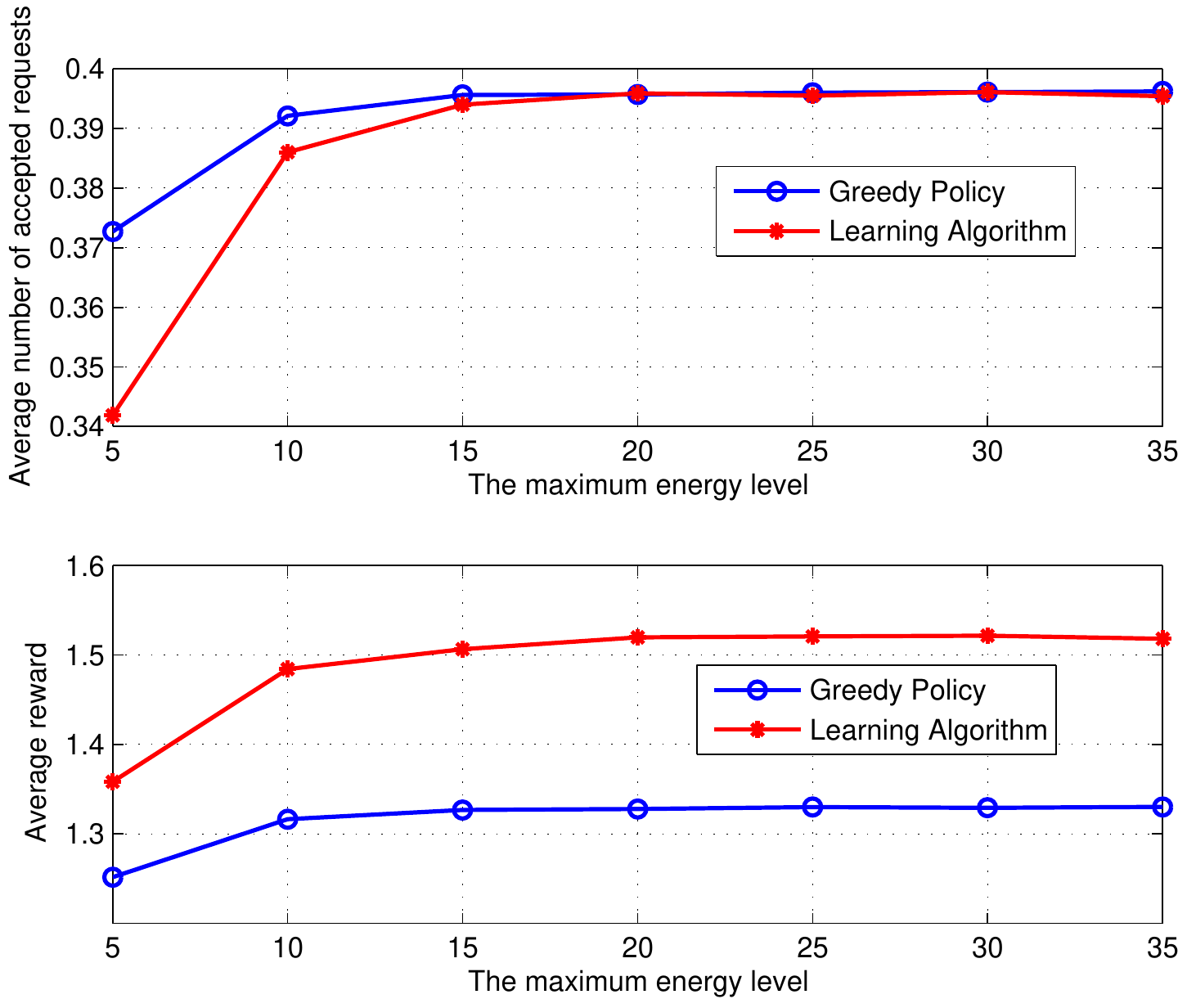}
\caption{The performance of the system when the maximum queue size is varied.}
\label{performance_of_system_vary_energy_queue}
\end{figure}
\begin{figure}[htb]
\centering
\includegraphics[scale=0.5]{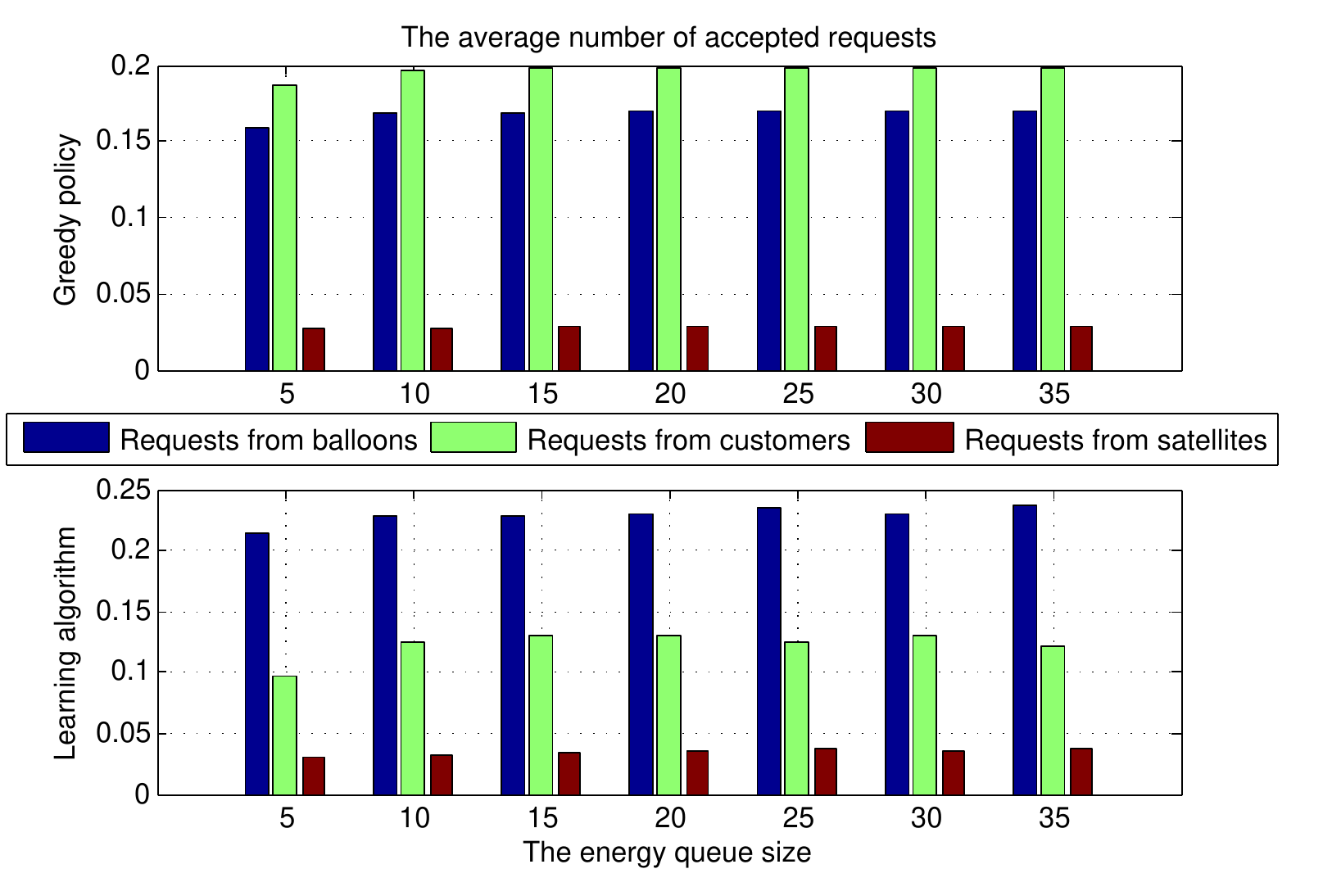}
\caption{The average number of accepted request.}
\label{average_no_accepted_requests}
\end{figure}

We then evaluate the impacts of the battery capacity to the performance of the system. Specifically, in Fig.~\ref{performance_of_system_vary_energy_queue}, we vary the maximum battery capacity and observe its impacts to the average number of accepted requests and the average reward of the access point. When the maximum battery capacity increases, the average reward and average number of accepted requests obtained by the learning algorithm (LA) and the greedy policy (GP) will increase and saturate when the maximum battery capacity is greater than $15$. However, it is interesting that when the maximum battery capacity increases from $5$ to $15$, the average number requests accepted by the learning algorithm is lower than that of the greedy policy. However, the average reward obtained by the learning algorithm is higher than that of the greedy policy. The reason can be found from the policy of the learning algorithm and the policy of the greedy policy as shown in Fig.~\ref{average_no_accepted_requests}. While the greedy policy always accepts requests if the battery is not empty, the learning algorithm selectively accepts requests from class-2 and class-3 when the energy level is high enough. It is also worth to note that, when the maximum battery capacity is greater than $15$, the average numbers of requests obtained by the learning algorithm and the greedy policy are equal. However, the average reward obtained by the learning algorithm is always greater than that of the greedy policy. The reason is because when the battery capacity is small, the amount of energy harvested will be limited and thus learning algorithm will accept requests which yield high reward and reject requests which yield low reward. When the battery capacity increases, the amount of energy harvested will increase, and thus, there is more chance for the requests with low reward to be accepted (as shown in Fig.~\ref{average_no_accepted_requests}). However, when the battery capacity is greater than $15$, the performance of the system will be saturated. The reason is, the number of accepted requests depends not only the battery capacity, but also on the energy arrival rate. In other words, the system performance is constrained by energy arrival, if the battery capacity is large enough.

\begin{figure}[h]
\centering
\includegraphics[height=65mm, width=170mm]{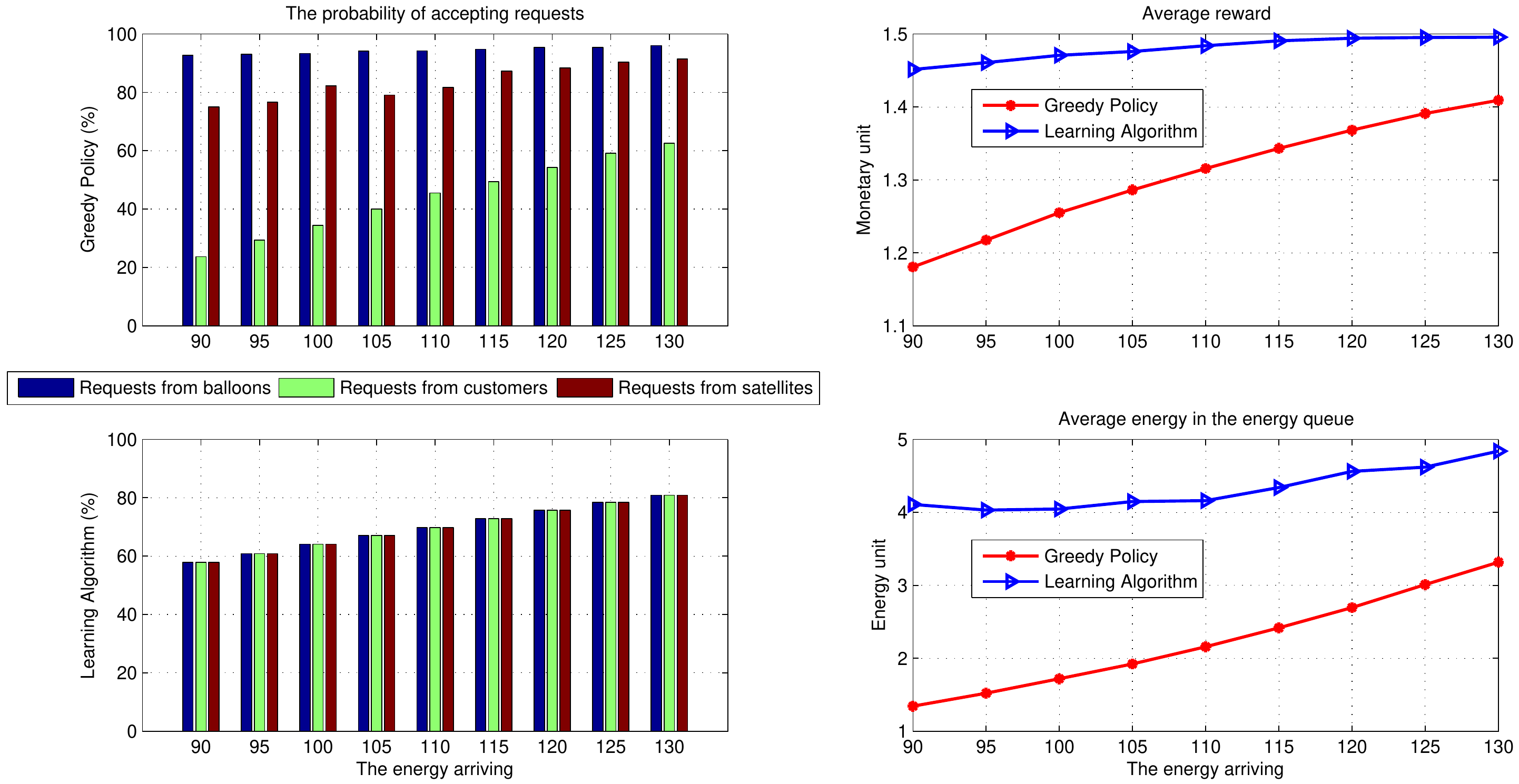}
\caption{The performance of the system when the energy arrival probability is varied.}
\label{energy_arriv_impact}
\end{figure}

We fix the battery capacity at $10$ and vary the energy arrival rate. When the energy arrival rate increases from $90$ to $130$, the probability of accepting requests by the greedy policy and the learning algorithm will increase. As a result, the average reward as well as the average energy harvested by both policies will also increase. Moreover, as shown in Fig.~\ref{energy_arriv_impact}, when the energy arrival rate is small, the average reward and the average energy in the battery with the learning algorithm are much greater than those of the greedy policy. However, when the energy arrival rate increases, the performance gap between the learning algorithm and greedy algorithm becomes smaller. Eventually, when the energy arrival rate is large, the results obtained by the greedy policy will approach those of the learning algorithm. 

\section{Summary} 
In this paper, we have studied and developed an optimization model for the optimal energy control problem for a network in the sky. The aim is to maximize the network performance as well as the profit for the network provider. We have first formulated the problem as a Markov decision process and then applied an online learning algorithm based on the gradient method to obtain the optimal policy for the access point deployed on a balloon. The numerical results have been presented to show the impacts of parameters to the system performance as well as to show the convergence and the efficiency of the proposed learning algorithm.

\bibliographystyle{IEEE}

\end{document}